\documentclass[runningheads]{llncs}
\usepackage{multirow}
\usepackage{amsmath,graphicx}
\begin{document}

\title{Deep Q-Network-Driven Catheter Segmentation in 3D US by Hybrid Constrained Semi-Supervised Learning and Dual-UNet}
\author{Hongxu Yang\inst{1}\and Caifeng Shan\inst{2} \and \\ Alexander F. Kolen\inst{3} \and
Peter H. N. de With\inst{1}} % 1{Yang, Hongxu}, 2{Shan, Caifeng}, 3{Kolen, Alexander F.}, 4{de With, Peter H. N}   
\institute{Eindhoven University of Technology, Eindhoven, The Netherlands\\ 
\and
Shandong University of Science and Technology, Qingdao, China
\and
Philips Research, Eindhoven, The Netherlands}
\titlerunning{DQN-Driven Catheter Segmentation by SSL}
\authorrunning{H. Yang et al.}
% corresponding author: Hongxu Yang, email: h.yang@tue.nl
\maketitle              
\begin{abstract}
Catheter segmentation in 3D ultrasound is important for computer-assisted cardiac intervention. However, a large amount of labeled images are required to train a successful deep convolutional neural network (CNN) to segment the catheter, which is expensive and time-consuming. In this paper, we propose a novel catheter segmentation approach, which requests fewer annotations than the supervised learning method, but nevertheless achieves better performance. Our scheme considers a deep Q learning as the pre-localization step, which avoids voxel-level annotation and which can efficiently localize the target catheter. With the detected catheter, patch-based Dual-UNet is applied to segment the catheter in 3D volumetric data. To train the Dual-UNet with limited labeled images and leverage information of unlabeled images, we propose a novel semi-supervised scheme, which exploits unlabeled images based on hybrid constraints from predictions. Experiments show the proposed scheme achieves a higher performance than state-of-the-art semi-supervised methods, while it demonstrates that our method is able to learn from large-scale unlabeled images.
\keywords{Catheter segmentation \and deep reinforcement learning \and Dual-UNet \and semi-supervised learning \and hybrid constraint.}
\end{abstract}
\section{Introduction}
Catheter segmentation in 3D ultrasound (US) images is of great importance in computer-assisted intervention, such as RF-ablation or cardiac TAVI procedures \cite{ICIP2019Yang,MICCAI2019Yang,IJCARS2019}. Catheter segmentation in 3D US is usually treated as a voxel-wise classification task, which assigns a semantic category to every 3D voxels (regions) by data-driven methods, such as using Deep Learning (DL) \cite{DLReview}. Typically, DL methods encounter challenges like limited training samples, expensive annotations and imbalanced classes. These factors degrade the DL performances for catheter segmentation in 3D US in different aspects  \cite{ICIP2018Yang}. To overcome the class-imbalance issue, a patch-based strategy was introduced to address the strongly imbalanced class distributions~\cite{ICIP2019Yang,MICCAI2019Yang}. However, this approach requires an accurate annotation for the training datasets, which is expensive for 3D US images and laborious for clinical experts. Moreover, a patch-based strategy requires the Convolutional Neural Network (CNN) to exhaustively segment the patches on the whole image, which is computationally expensive for real-time clinical usage. In the past years, several methods \cite{MICCAI2017Zhang,MICCAI2018Nie,BMVC2018,MASSL,MICCAI2019SSL1,MICCAI2019SSL2} have been proposed to also utilize the unlabeled images and thereby improve the segmentation performance for medical imaging. Adversarial learning has been studied to make use of unlabeled images, by either enforcing the network to perform segmentation on unlabeled images similar to the labeled images \cite{MICCAI2017Zhang}, or selecting the most reliable segmentations on unlabeled images to train the segmentation network \cite{MICCAI2018Nie}. Li \textit{et al.} \cite{BMVC2018} proposed to extend the $\Pi$-model for semi-supervised learning, which achieved promising results in skin lesion segmentation. A multi-task-based framework was proposed to exploit unlabeled images by attention-based image reconstruction \cite{MASSL}. An uncertainty-aware self-ensembling model was proposed \cite{MICCAI2019SSL1,MICCAI2019SSL2} to make use of certainty estimations for the segmentations of unlabeled images, which enhances the segmentation performance with limited annotations. Although uncertainty-aware methods \cite{MICCAI2019SSL1,MICCAI2019SSL2} achieved better performances than state-of-the-art methods (SOTAs), they are all based on the mean-teacher approach with exponential moving averaging on parameter updating, which encounters a parameter-correlation between teacher and student models \cite{DualStudent}. As a consequence, the reliability of used voxels may not be stable enough for SSL. 
\begin{figure}[tb]
\centering{\includegraphics[width=10cm]{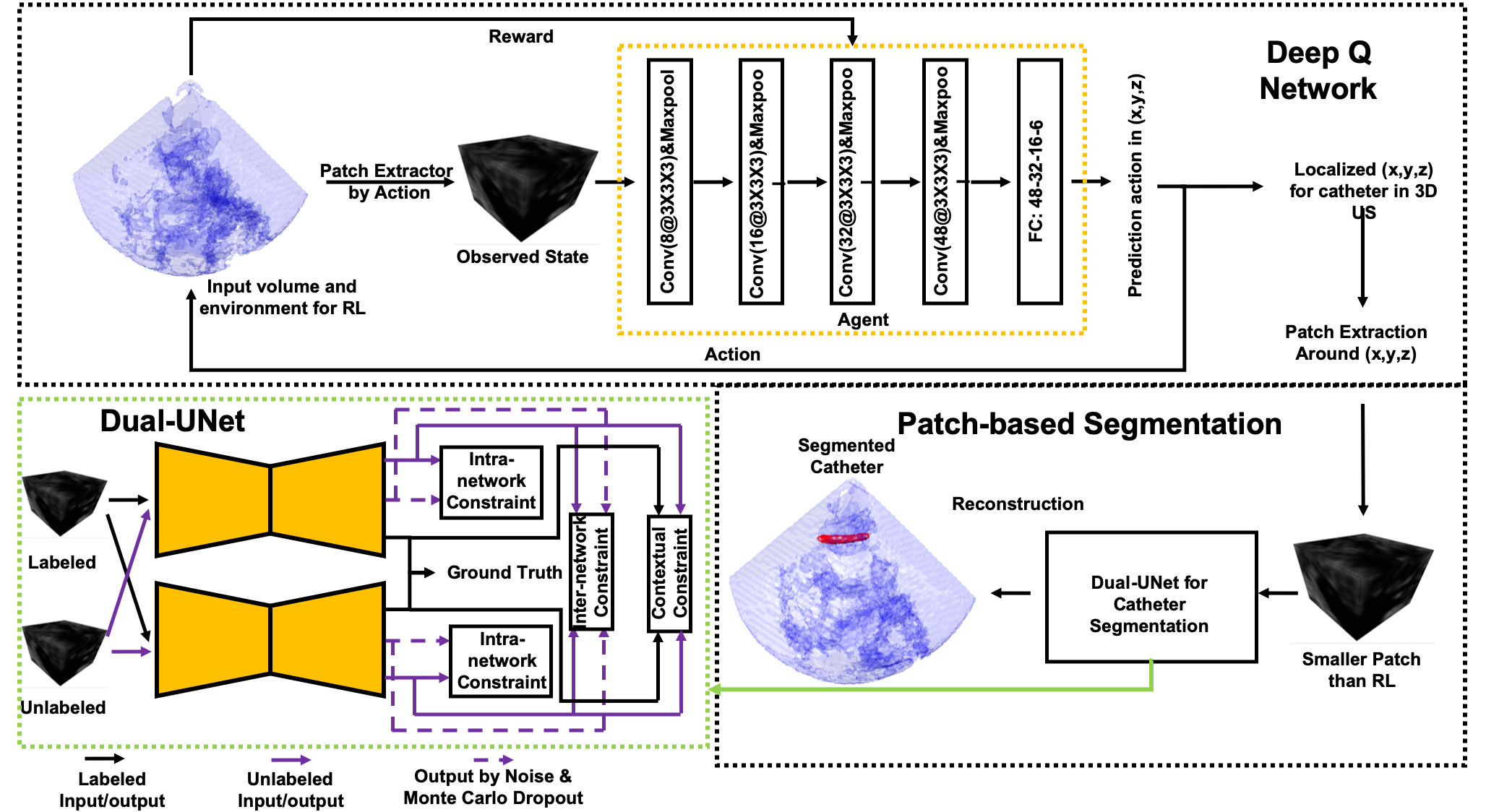}} % constraint for the figure.
\caption{Schematic view of the proposed framework.}
\label{overview}
\end{figure}

To address above challenges on annotation effort and segmentation strategies for patch-based CNN, we propose a deep Q learning-driven \cite{TPAMI2019}  catheter segmentation framework in 3D US by a semi-supervised approach that leverages reinforcement learning as a pre-localization step for catheter segmentation, which makes use of limited labeled and abundant unlabeled images to train a patch-based CNN. The proposed two-stage deep learning scheme is shown in Fig.~\ref{overview}. More specifically, the deep Q network (DQN) method performs coarse catheter localization by a learned search-path policy through the interaction with the 3D US image, which omits voxel-level annotations. Because of the DQN properties for discrete space actions, the annotation effort is drastically reduced, but the detection is degraded from a full voxel-level annotation to region-level semantic annotation. After the DQN-driven catheter localization, the proposed scheme introduces a Dual-UNet network, which segments the catheter around the localized region by a patch-based strategy \cite{ICIP2019Yang,MICCAI2019Yang}. To overcome the limited amount of labeled images and leverage the abundantly available unlabeled images, the proposed Dual-UNet is trained by a novel semi-supervised learning (SSL) scheme. In more detail, the proposed scheme consists of two independent UNets by learning the semantic information from a hybrid loss function, based on a supervised loss, intra-network uncertainty constraint loss, inter-network uncertainty constraint loss, and a contextual constraint loss. With uncertainty estimations within two UNets, the most reliable voxel-level predictions of unlabeled images are used as the semi-supervised signal to enhance the segmentation performance. To compensate the voxel-level constraints from the uncertainty estimation and enhance the usage of semantic information, we also introduce a contextual constraint, which enforces the network to learn the contextual information from unlabeled images. From our experiments, the proposed method was extensively compared to the SOTAs on an \textit{ex-vivo} dataset for cardiac catheterization. The results demonstrate that our method performs better than supervised learning method from same back-bone, while outperforming the SSL SOTAs.

Exploiting the proposed scheme for catheter segmentation, this paper presents the following contributions. We propose a novel catheter segmentation scheme based on the DNQ-driven Dual-UNet, which is trained by the proposed SSL approach. The proposed SSL training strategy is based on hybrid constraints and employs an unlabeled signal to improve the discriminating capacity of the CNN. The method leverages fewer annotations than a supervised learning approach, thereby reducing the training challenges. Moreover, the proposed scheme achieves a higher efficiency because of the use of a coarse-to-fine strategy.  
\section{Methods}
The proposed segmentation method is shown in Fig.~\ref{overview}. First, catheter center is detected by the DQN, which provides an estimation for the catheter location. Second, with the obtained catheter center, Dual-UNet is applied on local patches around the estimated location, which is trained by proposed SSL scheme. 
\subsection{Deep Q Learning as a pre-localization step for segmentation}
The task of extracting a 3D US region containing the catheter can be modeled under the reinforcement learning framework. The agent with its current observation state $s$ interacts with the environment $\mathcal{E}$, by performing successive actions $a\in{\mathcal{A}}$ to maximize the expected reward $r$. In this paper, we define the observation state in Cartesian coordinates as a 3D observation patch with size of $d^3$ voxels w.r.t. the patch center point $(x,y,z)$. To interact with $\mathcal{E}$, the action space $\mathcal{A}$ has six elements, which are defined as $\{\pm{a_x},\pm{a_y},\pm{a_z}\}_{scale}$ w.r.t. three different axes with a resolution of step \textit{scale}. With the observed state $s$, the agent makes a decision of an action from $\mathcal{A}$, to iteratively update the location of the 3D patch. After each action, a reward RL system is characterized as $r=sign(D(Pt_g,Pt_{t-1})/{scale}-D(Pt_g,Pt_{t})/{scale})$, where $D$ denotes the Euclidean distance between two points, $Pt_g$ is the ground-truth point, $Pt_t$ is the current state, $Pt_{t-1}$ is the previous state and $scale$ represents the step scale in $\mathcal{E}$ \cite{TPAMI2019}. As a result, reward $r\in\{-1,0,+1\}$ indicates whether the agent invokes a patch to move or leave it to the target point. With the obtained reward, the optimized action-selection policy can be characterized by learning a state-action value function $Q(s,a)$, which can be approximated by the Deep Q Network (DQN). To train the DQN, the loss function is defined as:
\begin{equation}\label{DQN}
\mathcal{L}_\text{DQN}=E_{s,r,a,\hat{a}\sim{M}}[(r+\gamma\cdot{\max}_{\hat{a}}Q(\hat{a},\hat{s};\widetilde{\omega})-Q(a,s;{\omega}))^2], 
\end{equation}
where the future reward discount parameter $\gamma$ is set to be 0.9, $\hat{a}$ and $\hat{s}$ are action and observed state in the next step, respectively. Parameter $M$ is the experience replay to de-correlate the random samples. Parameters $\omega$ and $\widetilde{\omega}$ are trainable parameters of the Q networks for current and target network, respectively. The architecture of the adapted Q network is depicted in Fig.~\ref{overview}, where four recent patches are used as the input. Search is terminated after local oscillation. 
\subsection{Semi-supervised catheter segmentation by Dual-UNet }
With coarse localization of the catheter in a 3D US volume, the catheter is then segmented by the proposed patch-based Dual-UNet, which is trained by a hybrid constrained framework. The proposed Dual-UNet structure is motivated by the mean-teacher architecture, which learns the network parameters by updating a student network from a teacher network \cite{MICCAI2019SSL1,MICCAI2019SSL2}. Intuitively, this method introduces two networks whose parameters are highly correlated due to the strategy of performing an exponential moving average on the updating process. As a result, the obtained knowledge is biased and may not be discriminative enough \cite{DualStudent}. Alternatively, we propose to use two independent networks, to learn the discriminating information by knowledge interaction through uncertainty constraints. Moreover, a contextual-level prediction constraint is introduced to compensate the information usage from voxel-level uncertainty estimations.

With the localized catheter from the DNQ, patches around it are extracted for semantic segmentation (or training). We studied the task of catheter segmentation in 3D US volumetric data, where training patches contain $N$ labeled patches and $M$ unlabeled patches. In this paper, labeled patches are denoted by $\{(x_i,y_i)\}_{i=1}^{N}$ and unlabeled patches are denoted as $\{x_j\}_{j=1}^{M}$, where $x\in{R}^{V^3}$ is the 3D input patch and $y\in\{0,1\}^{V^3}$ is the corresponding ground truth ($V=48$ as suggested by \cite{MICCAI2019Yang}). The task of semi-supervised learning is to minimize the following loss function with inputs of patches and ground truth (if applicable):
\begin{equation}\label{Loss}
\begin{split}
&\min_{\omega_1,\omega_2}(\sum_{i=1}^{N}(L_{sup}(f_1(x_i;\omega_1),y_i)+L_{sup}(f_2(x_i;\omega_2),y_i))\\
&+\alpha\sum_{j=1}^{M}(L_{intra}(f_1(x_j;\omega_1),f_1(x_j;\omega_1,\xi_1))+L_{intra}(f_2(x_j;\omega_2),f_2(x_j;\omega_2,\xi_2))\\
&+L_{inter}(f_1(x_j;\omega_1),f_2(x_j;\omega_2),f_1(x_j;\omega_1,\xi_1),f_2(x_j;\omega_2,\xi_2)))\\
&-\beta\sum_{i,j=1,1}^{N,M}L_{c}(f_1(x_i;\omega_1),f_2(x_i;\omega_2),f_1(x_j;\omega_1),f_2(x_j;\omega_2))),
\end{split}
\end{equation} 
where $\omega_1$, $\omega_2$ are trainable parameters, $L_{sup}$ denotes supervised loss (e.g. cross-entropy loss + Dice loss), $L_{intra}$ is the uncertainty constraint to measure the consistency between outputs from each individual network $f_q(\cdot)$ under different perturbations, where $q\in\{1,2\}$. Here, $\xi$ involves the dropout layers and Gaussian random noise of the input. Function $L_{inter}$ is an uncertainty constraint to measure the consistency between the two networks $f_1(\cdot)$ and $f_2(\cdot)$, which makes use of information from two independent networks to enhance the performance. Function $L_{c}$ is the cross-entropy loss w.r.t. the prediction with a label or without label for Adversarial learning \cite{MICCAI2017Zhang}. Coefficient $\alpha$ and $\beta$ were empirically selected as 0.1 and 0.002 to balance the losses.

\noindent {{\textbf {Uncertainty Intra-network Constraint $L_{intra}$:}}} Without annotation for an input patch, the prediction from networks to guide the unsupervised learning may be unreliable. To generate a reliable prediction from history and guide the network to gradually learn from the more reliable prediction, we design an uncertainty constraint for each individual network. Given an input patch, $T$ predictions are generated by a forward pass, based on a Monte Carlo Dropout and input with Gaussian noise \cite{MC}. Therefore, the estimated probability map is obtained by the average of $T$ times prediction for an input patch, i.e. $\hat{P_q}$ for network $f_q(\cdot)$. Based on the above probability maps, the uncertainty is obtained as $\hat{U_q}=-\hat{P_q}log(\hat{P_q})$ and the loss constraint for network $f_q(\cdot)$ is formulated by:
\begin{equation}\label{LC}
L_{intra}=\frac{\sum(\mathcal{I}(\hat{U_q}<\tau_1)\odot||f_q(x;\omega_q)-\hat{P_q}||)}{\sum\mathcal{I}(\hat{U_q}<\tau_1)},
\end{equation}
 where $\mathcal{I}$ is a binary indicator function, $\tau_1=-0.5ln(0.5)$ is a threshold to measure the uncertainty, which selects the most reliable voxels by binary voxel-level multiplication $\odot$. Parameter $f_q(x;\omega_q)$ is the prediction from network $f_q(\cdot)$ with $q\in\{1,2\}$. By following this approach, the proposed strategy is approximately equal to the mean-teacher method with the history step as unity.
  
\noindent {{\textbf {Uncertainty Inter-network Constraint $L_{inter}$:}}} Besides the above uncertainty constraint for each network, we also propose an uncertainty constraint to measure the consistency between two individual networks with the purpose to constrain the knowledge and avoid bias. The proposed uncertainty constraint on two networks can enhance the overall stability of the model by comparing the predictions between two networks. From the proposed networks in Fig.~\ref{overview} and the above definitions, different predictions can be obtained, expressed as $P_q$ and $\hat{P}_q$ for normal prediction and Bayesian prediction. Moreover, corresponding binary predictions are obtained as $C_q$ and $\hat{C}_q$, respectively, which are thresholded by 0.5. This leads to a stable prediction, which is formally written as
\begin{equation}\label{Stable}
\mathcal{S}_q=\mathcal{I}(C_q \odot\hat{C}_q)\odot(\mathcal{I}(U_q<\tau_2)\oplus\mathcal{I}(\hat{U_q}<\tau_2)),
\end{equation}
where $\tau_2=-0.7ln(0.7)$ is a threshold to select the more reliable voxels for $f_q(\cdot)$. Furthermore, we also define the voxel-level probability distance $D_q=||f_q(x;\omega_q)-\hat{P_q}||$, which indicates the network-wise consistency. Symbol $\oplus$ stands for a voxel-based logical OR. With the above, the uncertainty inter-network constraint $L_{inter}^1$ (where $L_{inter}=L_{inter}^1+L_{inter}^2$) for $f_1(\cdot)$ is formulated by:
\begin{equation}\label{LS}
L_{inter}^1=\frac{\sum(((\mathcal{S}_1\odot\mathcal{S}_2\odot\mathcal{I}(D_1>D_2))\oplus(\overline{\mathcal{S}_1\odot\mathcal{S}_2}\odot\mathcal{S}_2))\odot||P_1-P_2||)}{\sum((\mathcal{S}_1\odot\mathcal{S}_2\odot\mathcal{I}(D_1>D_2))\oplus(\overline{\mathcal{S}_1\odot\mathcal{S}_2}\odot\mathcal{S}_2))},
\end{equation}
where $||\cdot||$ is the probability distance at the voxel level and $\overline{(\cdot)}$ is binary NOT operation. This uncertainty constraint enables the unsupervised signal communication between two individual networks for $f_1(\cdot)$ and vice versa for $f_2(\cdot)$.

\noindent {{\textbf {Contextual Constraint $L_{c}$:}}} The above constraints only consider voxel-level consistency of paired predictions, while ignoring the differences between labeled and unlabeled predictions at the contextual level. To enhance the predictions similarity at contextual level, we also introduce a contextual constraint, which is based on the implementation of adversarial learning. The labeled and unlabeled predictions are classified by an adversarial classifier, which is learned by the loss $L_{c}$, i.e. BCE, based on whether it has annotation or not. Meanwhile, $L_{c}$ is maximized by Eqn.~(\ref{Loss}) to enhance the contextual similarity between predictions.
\section{Experimental Results}
\noindent {{\textbf {Materials and Implementation Details:}}} We collected 88 3D US volumes from 8 porcine hearts. During the data collection, the tissues were placed in water tanks with an RF-ablation catheter (diameter ranging from 2.3 mm to 3.3 mm) inside the chambers. The obtained images were re-sampled to have a volume size of $160\times160\times160$ voxels with padding applied, which leads to a voxel size ranging from $0.3^3$-$0.9^3$~mm. All the volumes were manually annotated at voxel level. Moreover, the catheter centers were also marked as the target location for DQN. To validate proposed method, 60 volumes were randomly selected as training set and 28 volumes were used as testing images. To train the DQN, 60 volumes with target location were used to learn the path-search policy. To train the Dual-UNet, 10\%, 20\% and 30\% of 60 images were selected as the labeled images, while the remainder were the unlabeled images for SSL. 
\begin{figure}[tb]
\centering{\includegraphics[width=10cm]{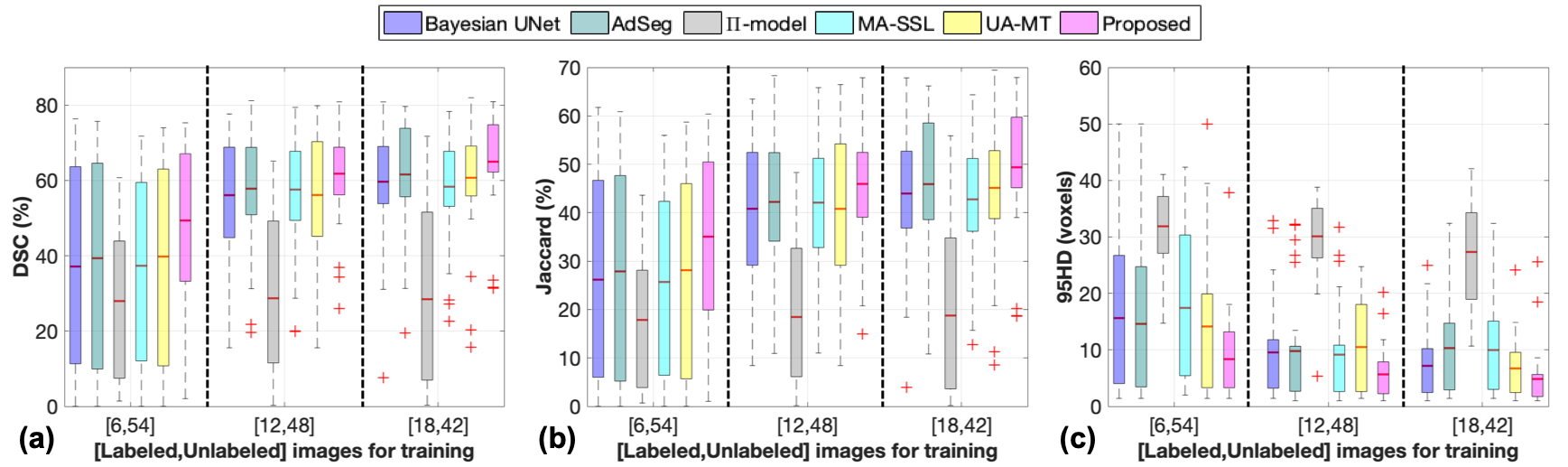}} % constraint for the figure.
\caption{Comparison to SOTAs by boxplot: (a) Dice Score, (b) Jaccard Score, (c) 95HD. Higher DSC/Jaccard and lower 95HD show better results for segmentation. }
\label{comparison}
\end{figure}

\begin{table}[tb]
\centering
\caption{Segmentation performance by ablation studies in Dice Score (DSC), Jaccard Score (Jaccard) and 95\% Hausdorff Distance (95HD), which are shown in mean$\pm$std.}
\label{Compare}	
\begin{tabular}{l|c|c|c|c|c}
\hline
\multirow{2}{*}{Method}& \multicolumn{2}{c|}{\emph{\# Images Used}}& \multicolumn{3}{c}{\emph{Metrics}}\\ \cline{2-6} 
& Labeled &Unlabeled& DSC (\%)$\uparrow$& Jaccard (\%)$\uparrow$&95HD (voxels)$\downarrow$\\ \hline
DU-$L_{intra}$&6&54&42.4$\pm$25.1&29.8$\pm$19.2&13.9$\pm$11.2\\\hline
DU-$L_{intra+inter}$&6&54&{47.9$\pm$23.5}&{34.2$\pm$19.4}&{11.7$\pm$8.3}\\\hline
\textbf{Proposed}&6&54&\textbf{49.4$\pm$20.7}&\textbf{35.1$\pm$17.3}&\textbf{8.5$\pm$7.8}\\\hline\hline
DU-$L_{intra}$&12&48&58.8$\pm$17.0&43.5$\pm$15.9&8.2$\pm$9.4\\\hline
DU-$L_{intra+inter}$&12&48&{59.6$\pm$13.2}&{43.7$\pm$13.1}&{7.8$\pm$8.5}\\\hline
\textbf{Proposed}&12&48&\textbf{61.8$\pm$13.4}&\textbf{45.9$\pm$13.3}&\textbf{5.7$\pm$4.6}\\\hline\hline
DU-$L_{intra}$&18&42&62.6$\pm$14.0&46.8$\pm$13.4&6.2$\pm$6.6\\\hline
DU-$L_{intra+inter}$&18&42&{63.8$\pm$13.3}&{48.1$\pm$13.3}&{5.2$\pm$5.9}\\\hline
\textbf{Proposed}&18&42&\textbf{65.0$\pm$13.5}&\textbf{49.4$\pm$13.3}&\textbf{4.7$\pm$5.1}\\\hline\hline
UNet\cite{MICCAI2019Yang}&60&0&63.0$\pm$14.4&47.3$\pm$13.2&5.5$\pm$7.1\\\hline\hline
\end{tabular}
\end{table}

We implemented our framework in \textit{TensorFlow}, using a standard PC with a TITAN 1080Ti GPU. We trained the DQN with Adam \cite{Adam} optimizer (learning rate 1e-4) for 40 epochs. Replay memory was 1e5. Parameters of the target network were updated for every 2,500 steps. When considering the efficiency and accuracy of the DQN, we defined the input state space to $55^3$ voxels for resized images with size $128^3$, which promises the observations can contain sufficient catheter contextual information. To construct the Dual-UNet as the segmentation network, we employ Compact-UNet \cite{MICCAI2019Yang} as our backbone, together with a joint cross-entropy and Dice loss. To adapt the UNet as a Bayesian network and generate uncertainty prediction, dropout layers with rate 0.5 were inserted prior to the convolutional layers. Gaussian random noise was also considered during uncertainty estimation. For the uncertainty estimation, $T=8$ was used to balance the efficiency and quality of the estimation. As for $L_{c}$, the classifier was constructed based on the contextual encoder from \cite{MICCAI2019Yang} with fully connected layers 128-32-1. Training was terminated after 10,000 steps with mini-batch size of 4 using the Adam optimizer, which includes 2 labeled and 2 unlabeled patches. The ramp-up weight was applied on $\tau_1$ and $\tau_2$ to control the training process.

\noindent {{\textbf {Results Analysis:}}} As for the experiments of using DQN as pre-selection, we compared localization accuracy for different volume sizes as environment: $128^3$ and $160^3$ voxels, since this would use different amounts of contextual information within the fixed observation space. The metric is the Euclidean distance between the detected catheter center point and ground-truth center point by voxels in the resolution of $160^3$ voxels. Statistical performance of the $128^3$ case is {$4.3\pm2.6$} voxels, while it becomes $4.7\pm4.8$ voxels for the $160^3$ voxels case (measured by mean $\pm$ std.). From the result, $128^3$ voxels' environment provides a better detection accuracy without failure because of the increased contextual information that is observed by the agent for a fixed input size. 

With the detected catheter center point, patches with size of $48^3$ are extracted around the point for semantic segmentation (i.e. $2^3$ patches). Segmentation results are obtained based on DQN detection. To evaluate the performance, we consider the Dice Score (DSC), Jaccard Score and 95\% Hausdorff Distance (95HD) as the evaluation metrics \cite{Metric}. We have implemented several state-of-the-art (SOTA) SSL methods for comparison, which include Bayesian UNet \cite{MC}, $\Pi$-model \cite{BMVC2018} Adversarial-based segmentation (AdSeg)\cite{MICCAI2017Zhang}, multi-task attention-based SSL (MA-SSL)\cite{MASSL} and uncertainty-aware-based mean-teacher (UA-MT)\cite{MICCAI2019SSL2} (all of them are based on Compact-UNet \cite{MICCAI2019Yang}). Results are shown in Fig.~\ref{comparison}, which shows the proposed method outperforms SOTA SSL approaches by at least 4\% DSC on average. As the number of labeled images increases, the segmentation performances are improved w.r.t available supervised information except for the $\Pi$-model, which is due to the unreliable information of the predicted unlabeled images. Because of the unreliable information in challenging 3D US images, the $\Pi$-model obtains a lower performance than a simple Bayesian UNet. Compared to MA-SSL and UA-MT, our method achieves a better performance, since it exploits inter/intra-network uncertainty information and boosts the information usage of unlabeled images. To perform ablation studies, the loss components in Eqn.~(\ref{Loss}) are gradually introduced based on the Dual-UNet (denoted as DU), which are shown in Table.~\ref{Compare}. The ablation studies show the proposed constraint component can make use of unlabeled information at different levels, and therefore improve the performances. As can be observed, using 18 labeled images and 42 unlabeled images, the proposed scheme achieves a higher performance than the supervised learning method. From the experiments, the proposed two-stage scheme achieves around 1.2 sec. per volume (0.7+0.5 sec.). As a comparison, exhaustive patch-based segmentation spends $>10$ sec. per volume \cite{MICCAI2019Yang}, while a voxel-of-interest-based CNN method costs around 10 sec.~\cite{IJCARS2019}. 
\section{Conclusion}
In this paper, we have proposed a catheter segmentation scheme using a DQN-driven semi-supervised learning for US-guided cardiac intervention therapy. In the proposed method, we design a DQN to coarsely localize the catheter in the 3D US, which is then segmented by an SSL-trained Dual-UNet. With extensive comparison, the proposed method outperforms the SOTAs, while it achieves a higher performance than the supervised learning approach with fewer annotations. Future work will investigate a more complex model with a better supervised loss but to obtain a higher efficiency.
\bibliographystyle{unsrt}

\end{document}